\documentclass[prd,aps,showpacs,nofootinbib,showkeywords,eqsecnum]{revtex4-1}

\usepackage{graphicx,color,amsmath,amsxtra}
\usepackage{epsf}
\usepackage{amssymb}
\usepackage{enumerate}
\usepackage{hhline}
\usepackage{array}
\usepackage{tabularx}
\usepackage{caption}
\usepackage[unicode]{hyperref}
\usepackage{graphicx}                
\usepackage{epstopdf}

\begin{document}

\pagestyle{myheadings}


\title{Anisotropic power-law inflation for a conformal-violating Maxwell model}
\author{Tuan Q. Do }
\email{tuanqdo@vnu.edu.vn}
\affiliation{Faculty of Physics, VNU University of Science, Vietnam National University, Hanoi 120000, Vietnam}
\author{W. F. Kao}
\email{gore@mail.nctu.edu.tw}
\affiliation{Institute of Physics, Chiao Tung University, Hsin Chu 30010, Taiwan}
\date{\today } 


\begin{abstract}
A set of power-law solutions of a conformal-violating Maxwell model with a non-standard scalar-vector coupling will be shown in this paper. In particular, we are interested in a coupling term of the form $X^{2n} F^{\mu\nu}F_{\mu\nu}$ with $X$ denoting the kinetic term of the scalar field. Stability analysis indicates that the new set of anisotropic power-law solutions is unstable during the inflationary phase. The result is consistent with the cosmic no-hair conjecture. We show, however, that a set of stable slowly expanding solutions does exist for a small range of parameters $\lambda$ and $n$. Hence a small anisotropy can survive during the slowly expanding phase.

Keywords: expanding universes, Bianchi spaces, no-hair theorem.
\end{abstract}


\keywords{higher derivative modes, expanding universes, Bianchi spaces}

\pacs{95.30.Sf, 98.80.Jk, 04.50.Kd, 98.80.Bp}

\maketitle



\section{Introduction}\label{sec1}
The cosmic inflation \cite{guth} has been regarded as a main paradigm of the modern cosmology. 
It follows from the success of the cosmic inflation resolving quite a number of cosmological problems such as the {\it horizon}, {flatness}, and {\it magnetic-monopole} problems ~ \cite{guth}.  
In addition, the standard  inflation model with our early universe assumed to be homogeneous and isotropic \cite{FLRW}  has been important in realizing many cosmic microwave background (CMB) observation results inclusign the Wilkinson Microwave Anisotropy Probe (WMAP) ~ \cite{Komatsu:2010fb} and the Planck~\cite{Planck}.  

Some recent CMB anomalies was, however, found. 
For example the hemispherical asymmetry and the cold spot have been detected by the WMAP and Planck.
Anisotropic models have therefore been considered as a genearalization of the FLRW inflation \cite{FLRW}. 
Consequently, modifications for the FLRW inflation are also necessary in order to accommodate the nature of the observed anomalies of CMB. 
It turns out that one of the simplest modifications is by replacing the FLRW metric by anisotropic but homogeneous Bianchi type spacetimes \cite{bianchi}. 
Note that many predictions made with anisotropic inflation have been done before the mentioned anomalies were detected~ \cite{Pitrou:2008gk}. 

The Bianchi models have also been discussed extensively in providing evidences for the  cosmic no-hair conjecture proposed by Hawking and his colleagues \cite{GH}. 
The cosmic no-hair conjecture postulates that the final-time state of our universe will be simply homogeneous and isotropic regardless of any initial conditions and states  of the universe in early time \cite{GH}.  
This  conjecture has been a great challenge to the physicists and cosmologists for several decades. 
Some partial proofs to this conjecture have been worked out. 
In particular, this conjecture has been theoretically proved for the Bianchi spaces by Wald with an energy conditions approach \cite{wald}.  

Recently, some people have tried to extend the Wald's proof to a more general scenario, in which the spacetime of universe is assumed to be inhomogeneous \cite{inhomogeneous}. Note that some other interesting proofs for this cosmic no-hair conjecture can be seen in  Refs. \cite{Barrow:1987ia,Carroll:2017kjo}. Besides the theoretical approaches, a general test using Planck’s data on the CMB temperature and polarization has provided  observational constraints on the isotropy of the universe \cite{Saadeh:2016sak}, by which we might see how precise the cosmic no-hair conjecture is.

The validity of the cosmic no-hair conjecture has also been tested in various cosmological models, including the higher curvature models ~\cite{barrow06,kao09}, the Lorentz Chern-Simons theory ~\cite{kaloper}, the massive vector theories ~\cite{Kanno:2008gn},  the nonlinear massive gravity models~\cite{massive-gravity}, the massive bigravity ~\cite{bigravity}, and the supergravity-motivated models ~\cite{MW,SD,WFK,WFK1,extensions,correlations,gravitational-waves}. 
Among these models, an interesting counter-example to the cosmic no-hair conjecture arising in the supergravity-motivated model proposed by Kanno, Soda, and Watanabe (KSW)~\cite{MW} has attracted many attentions \cite{SD,WFK,WFK1,extensions,correlations,gravitational-waves}. 

As a result, the KSW model  has been shown to admit a stable and attractive Bianchi type I inflationary solution \cite{MW} due to the existence of an unusual coupling term between scalar field and electromagnetic field, $f^2(\phi)F_{\mu\nu}F^{\mu\nu}$. 
More interestingly, non-canonical extensions of the KSW model, in which a scalar field takes non-canonical form such as Dirac-Born-Infeld (DBI), supersymmetric Dirac-Born-Infeld (SDBI), and covariant Galileon forms, have also been shown to admit stable and attractive Bianchi type I inflationary solutions \cite{WFK1}. 
This indicates that the cosmic no-hair conjecture does not hold for the extended KSW models due to the presence of the coupling term $f^2(\phi)F_{\mu\nu}F^{\mu\nu}$.

Note that the KSW model can be regarded as a subclass of the conformal-violating Maxwell theory with an extended coupling term $I\left({\phi,R,X,\ldots}\right)F_{\mu \nu } F^{\mu \nu } $~\cite{ITD,MST,BR,DMR,Grasso:2000wj,Barrow:2011ic}. 
Here $I\left({\phi, R,X,\ldots}\right)$ is a function of any field of interest. 
It is known that the conformal invariance must be broken to generate non-trivial magnetic fields~\cite{ITD,MST,BR,DMR,Grasso:2000wj,Barrow:2011ic}. 
Indeed, the coupling between scalar and vector fields, $\exp[\phi]F_{\mu \nu } F^{\mu \nu }$, has been proposed in Ref.~\cite{BR}  as a natural origin of  large-scale galactic electromagnetic field in the present universe.
 
Hence a close relation exists between the stable spatial anisotropy of inflationary universe and the broken conformal invariance. 
Indeed, the  mechanism of the broken conformal invariance should induce both non-trivial magnetic fields and a stable spatial anisotropy of spacetime during an inflationary phase.  
One might also consider some subclasses of the conformal-violating Maxwell theory to see whether the cosmic no-hair conjecture breaks down. 
In particular, a model with a Ricci scalar  non-minimally coupled to the electromagnetic field via a coupling $Y(R)F_{\mu\nu}F^{\mu\nu}$ \cite{Adak:2016led} has already been considered. 

In this paper, we will focus on a possible coupling term with $I=J^2\left({X}\right)$ as an algebraic function of the scalar field kinetic term  $X\equiv-\partial^\mu \phi \partial_\mu \phi /2$. 
Note that $J$ is considered as an arbitrary function of $X$. 
As a result, we will show that a set of Bianchi type I expanding power-law solutions exists. 
It will also be shown that these solutions are unstable during the inflationary phase in consistent with the cosmic no-hair conjecture in this model. 
On the other hand, we can also show that stable solutions do exist during the slowly-expanding phase.

This paper will be organized as follows: (i) A brief review and the motivation of this research have been given in Sec. \ref{sec1}. (ii) A complete setup of the proposed model will be presented in Sec. \ref{sec2}. (iii) A set of Bianchi type I power-law  inflationary solutions and its stability will be solved and discussed in Sec. \ref{sec3} and Sec. \ref{sec4}, respectively. (iv) Finally, concluding remarks will be given in Sec. \ref{sec5}.
\section{Conformal-violating Maxwell model} \label{sec2}
A conformal-violating Maxwell theory can be described by the general action given by ~\cite{DMR}
\begin{equation} \label{action}
S = \int {d^4 } x\sqrt {- g} \left[ {\frac{{R}}
{2} +X- V\left( \phi  \right) - \frac{1}
{4}I\left({\phi, R,X,\ldots}\right)F_{\mu \nu } F^{\mu \nu } } \right],
\end{equation}
with $F_{\mu \nu }  \equiv \partial _\mu  A_\nu   - \partial _\nu  A_\mu  $  the  field strength of the vector field $A_\mu$ and  $I\left({\phi, R,X,\ldots}\right)$ a function of any field of interest. 
For example, the models with $I= I\left(\phi\right)$ ~\cite{MW,SD,WFK,WFK1,extensions,correlations,gravitational-waves,BR},  $ I= I\left(R,R_{\mu\nu},R_{\mu\nu\lambda\kappa}\right)$ ~\cite{ITD,MST,Adak:2016led,FDM}, $I=I\left({A^2}\right)$ ~\cite{AG}, $I=I\left({G}\right)$ ~\cite{MRS,GEFC}, and $I=I\left[{(k_F)_{\alpha\beta\mu\nu}}\right]$ ~\cite{LCVAK} have been considered extensively in the literature. 
Note that the Planck mass $M_p$ has been set as one for convenience. 
In this paper, we will study a conformal violating model with $ I=J^2\left(X\right)$:
\begin{equation} \label{action1}
S = \int {d^4 } x\sqrt {- g} \left[ {\frac{{R}}
{2} +X- V\left( \phi  \right) - \frac{1}
{4}J^2\left(X\right)F_{\mu \nu } F^{\mu \nu } } \right].
\end{equation}
As a result, the field equations can be shown to be
\begin{eqnarray}
 \label{eq1}
\partial_\mu \left[{\sqrt {- g}J^2 F^{\mu\nu}}\right] &=&0, \\
 \label{eq2}
\left({1-\frac{1}{2}JJ' F^2}\right)\square \phi  -\frac{1}{2}\left({J J''+J'^2}\right)\partial_\mu X \partial^\mu \phi F^2   -\frac{1}{2}J J' \partial^{\mu}\phi \partial_{\mu}F^2-\partial_\phi V&=&0, \\
 \label{eq3}
G_{\mu\nu}-\left({1-\frac{1}{2}JJ' F^2}\right)\partial_\mu \phi \partial_\nu \phi+g_{\mu\nu}\left({\frac{1}{2}\partial_\sigma \phi \partial^\sigma \phi+V+\frac{1}{4}J^2F^2}\right) 
-J^2F_{\mu\gamma}F_\nu{}^\gamma &=&0. 
\end{eqnarray}
Here $J' \equiv \partial_X J $ denotes the differentiation with respect to the argument of $J(X)$.
In addition, the Einstein tensor is defined as $G_{\mu\nu} \equiv {R_{\mu\nu}-\frac{1}{2}R g_{\mu\nu}}$.
Similar to Refs. ~\cite{MW,WFK,WFK1}, we will focus on the Bianchi type I metric given by
\begin{equation} \label{eq4}
ds^2  =  - dt^2  + \exp \left[ {2\alpha \left( t \right) - 4\sigma \left( t \right)} \right]dx^2  
+ \exp \left[ {2\alpha \left( t \right) + 2\sigma \left( t \right)} \right]\left( {dy^2  + dz^2 } \right) .
\end{equation}
In addition, the scalar field $\phi$ and the vector field $A_\mu$ will be chosen as $\phi  = \phi \left( t \right)$ and $A_\mu   = \left( {0,A_x \left( t \right),0,0} \right)$ respectively. 
As a result,  Eq. (\ref{eq1}) can be integrated directly as
\begin{equation} \label{eq5}
\dot A_x\left({t}\right)=J^{-2}\left({X}\right)\exp\left[{-\alpha-4\sigma}\right]p_A,
\end{equation}
with $p_A$ the constant of integration ~\cite{MW}.
Hence we can rewrite Eq. (\ref{eq2}) as
\begin{equation} \label{eqnphi}
\left\{{1+J^{-3}\left[{J'+\dot\phi^2\left({J''-3J^{-1}J'^2}\right)}\right]\exp\left[{-4\alpha-4\sigma}\right]p_A^2}\right\}\ddot\phi = -3\dot\alpha \dot\phi+ \left({\dot\alpha+4\dot\sigma}\right)\dot\phi J^{-3}J' \exp\left[{-4\alpha-4\sigma}\right]p_A^2-\partial_\phi V. 
\end{equation}
In addition, the non-vanishing components of the Einstein equation (\ref{eq3}) can be shown to be:
\begin{eqnarray}
  \label{friedmann}
\dot \alpha ^2 & =& \dot \sigma ^2 +\frac{1}{3}\left[{\frac{\dot\phi^2}{2}+V+\frac{J^{ - 2}}
{2}\left({2\dot\phi^2 J^{-1}J'+1}\right) \exp \left[{ - 4\alpha  - 4\sigma }\right] p_A^2}\right] ,
\\  \label{eqnalpha}
\ddot \alpha & =&  - 3\dot \alpha ^2  +V+\frac{J^{-2}}{6}\left({3\dot\phi^2 J^{-1}J' +1}\right)\exp \left[{ - 4\alpha  - 4\sigma }\right] p_A^2 ,
\\ \label{eqsigma}
\ddot \sigma  &=&  - 3\dot \alpha \dot \sigma  + \frac{{J^{ - 2} }}
{3}\exp \left[{ - 4\alpha  - 4\sigma }\right] p_A^2 .
\end{eqnarray}
Note that the Friedmann equation \eqref{friedmann} is the $G^0_{~0}$ component equation, Eq. \eqref{eqnalpha} is the $(3G^0_{~0}+G^1_{~1}+2G^2_{~2})$ equation, and Eq. \eqref{eqsigma} is the $\left(G^1_{~1}-G^2_{~2}\right)$ equation. 
\section{Anisotropic power-law solutions} \label{sec3}
We will try to find a set of power-law solutions for the proposed model \eqref{action} in this section. In particular, we would like to obtain a new set of power-law analytic solutions with the form~\cite{MW,WFK,WFK1}:
\begin{equation} \label{eq6}
\alpha = \zeta \log \left( t \right);~ \sigma  = \eta \log \left( t \right);~ {\phi }
  = \xi \log \left( t \right) + \phi _0.
\end{equation}
Note that this set of power-law ansatz is the key to obtain a consistent set of power-law solutions.
For simplicity,  we will focus on the exponential scalar potential given by:
\begin{equation} \label{eq7}
V\left(\phi\right) = V_{0} \exp \left[{ \lambda \phi }\right],
\end{equation}
along with the power-law kinetic function:
\begin{equation} \label{eq8}
J\left({X}\right)=J_0 X^n.
\end{equation}
Here $V_0$, $J_0$, $\phi_0$, and $\lambda$ are constants specifying the boundary information of these potential and kinetic functions. 
In addition, the constants $\lambda$ and $n$ are assumed to be positive. 
For convenience, we will also introduce the following new variables:
\begin{eqnarray}
u &=& {{V_{0} \exp \left[{\lambda \phi _0 }\right] }},\\  
v &= &p_A^2 J_0^{ - 2}\left({\frac{\xi^2}{2}}\right)^{-2n}.
\end{eqnarray}
As a result, we can derive the following set of algebraic equations from the field equations:
\begin{eqnarray}
\label{eq9}
\left[{\frac{2n}{\xi^2}\left({4n+1}\right)v-1}\right]\xi +3\zeta\xi - \frac{2n}{\xi}\left({\zeta+4\eta}\right)v+\lambda u&=&0, \\
\label{eq10}
\zeta^2-\eta^2-\frac{\xi^2}{6}-\frac{u}{3}-\frac{1}{6}\left({4n+1}\right)v&=&0, \\
 \label{eq11}
-\zeta+3\zeta^2-u-\frac{1}{6}\left({6n+1}\right)v&=&0, \\
 \label{eq12}
\left({3\zeta-1}\right)\eta-\frac{v}{3}&=&0,
\end{eqnarray}
along with the following constraints:
\begin{eqnarray}
\label{eq13}
\lambda \xi &=&-2, \\
\label{eq14}
\zeta+\eta-n-\frac{1}{2}&=&0.
\end{eqnarray}
The constraint (\ref{eq14}) indicates that $n$ behaves similar to the ratio $\rho/\lambda$ discussed in the KSW model in Ref. ~\cite{MW} with $I=f^2(\phi) \sim \exp[2\rho \phi]$ and $V(\phi) \sim \exp[\lambda \phi]$. 
This is, however, the only similarity between the KSW model ~\cite{MW} and our proposed model. 
Indeed, it turns out that all field equations in this paper are not alike as compared to the field equations of the KSW model. 
With the constraint (\ref{eq13}), the variable $v$ can be written as
\begin{equation} \label{eq15}
v = p_A^2 J_0^{ - 2}\left({\frac{\lambda^2}{2}}\right)^{2n}.
\end{equation}
Note that both variables, $u$ and $v$, are assumed to be positive. 
Given $\eta$ shown in Eq. (\ref{eq14}), Eqs. (\ref{eq11}) and (\ref{eq12}) can be solved to give:
\begin{eqnarray}
&& \label{eq16}
u=\frac{1}{4}\left({3\zeta-1}\right)\left[{\left({6n+1}\right)\left({2\zeta-2n-1}\right)+4\zeta}\right], \\
&& \label{eq17}
v= -\frac{3}{2}\left({3\zeta-1}\right)\left({2\zeta-2n-1}\right).
\end{eqnarray}
With $\eta$, $u$, and $v$ defined above, scalar field equation (\ref{eq9})  and  Friedmann equation (\ref{eq10}) can be reduced as
\begin{eqnarray}
\label{eq18}
\left({3\zeta-1}\right)  \left[{36n\zeta^2-6\left({6n^2+3n-1}\right)\zeta-2n-8 \lambda^{-2}-1}\right ] &=&0, \\
 \label{eq19}
36n\zeta^2-6\left({6n^2+3n-1}\right)\zeta-2n-8 \lambda^{-2}-1&=&0,
\end{eqnarray} 
respectively. 
It turns out that both   Eqs. (\ref{eq18})  and (\ref{eq19})  lead to the non-trivial solutions
\begin{eqnarray} \label{eq20}
\zeta_{\pm}&=&\frac{6n^2+3n-1}{12n}\pm\frac{\sqrt{\Delta}}{12\lambda n},\\
\Delta&=&\lambda^2\left({36n^4+36n^3+5n^2-2n+1}\right)+32n 
\end{eqnarray}
for $\zeta \ne 1/3$.
Since $n$ is assumed to be positive and $\zeta > \eta$,  $\zeta_{+}$ can be shown to be only the consistent solution:
\begin{equation}
\zeta =\zeta_{+}=\frac{6n^2+3n-1}{12n}+\frac{\sqrt{\lambda^2\left({36n^4+36n^3+5n^2-2n+1}\right)+32n}}{12\lambda n}.
\end{equation}
In addition, the variable $\eta$ becomes
\begin{equation} \label{sol.eta}
\eta=\frac{6n^2+3n+1}{12n}- \frac{\sqrt{\lambda^2\left({36n^4+36n^3+5n^2-2n+1}\right)+32n}}{12\lambda n}.
\end{equation}
It is clear that $\eta <\zeta=\zeta_+$ as expected.
Following  Ref.~\cite{MW}, the anisotropy parameter is given by 
\begin{equation}
\frac{\Sigma}{H}\equiv \frac{\dot\sigma}{\dot\alpha}=\frac{(6n^2+3n+1)\lambda- \sqrt{\lambda^2\left({36n^4+36n^3+5n^2-2n+1}\right)+32n}}{(6n^2+3n-1)\lambda+\sqrt{\lambda^2\left({36n^4+36n^3+5n^2-2n+1}\right)+32n}}.
\end{equation}
The positivity of $v$ implies that $\eta >0 $ with the help of Eq. (\ref{eq12}) provided that $\zeta >1/3$. 
Hence, the positivity of $\eta$ leads to a constraint of $n$ given by
\begin{equation} \label{inequality}
n > \frac{4-\lambda^2}{2\lambda^2} =\frac{2}{\lambda^2}-\frac{1}{2},
\end{equation}
with the help of Eq. (\ref{sol.eta}).
Note also that for an anisotropically expanding solution, the following constraints $\zeta+\eta >0$ and $\zeta-2\eta >0$ need to be satisfied. 
It is apparent that the constraint $\zeta+\eta >0$ holds for positive $n$. 
In addition, the constraint $\zeta-2\eta >0$ leads to the inequality:
\begin{equation}\label{inequality2}
\lambda^2\left({8 n^3+8n^2-1}\right) +8>0.
\end{equation}
The inequality \eqref{inequality2} holds apparently for all $n>(\sqrt{5}-1)/4\approx 0.309$ if $\lambda \ge 0$. 
Hence, expanding solutions exist for the canonical model with $n=1$.
For  $0<n<(\sqrt{5}-1)/4$, the above inequality will depend, however, on the value of $\lambda$. 
Note also that the parameter $\Delta$ is positive definite if the inequality (\ref{inequality2}) holds. 
Finally, with the help of Eq. (\ref{eq16}), the positivity of $u$ leads to the inequality
\begin{equation}
\left({6n+1}\right)\left({2\zeta-2n-1}\right)+4\zeta >0,
\end{equation}
or equivalently, 
\begin{equation}\label{inequality3}
0<n<\zeta-\frac{1}{6}.
\end{equation}
Consequently, the following constraint on $\eta$ can be obtained from Eq. (\ref{eq14}) as 
\begin{equation}
\eta<\frac{1}{3}.
\end{equation}
It appears that there are constraints of $n$ given by the Eqs. (\ref{inequality}), (\ref{inequality2}), and (\ref{inequality3}) for the expanding solution with $\zeta+\eta >0$ and $\zeta-2\eta >0$. 
Note that we are interested in the inflationary phase solutions with the following constraints $\zeta+\eta \gg1$ and $\zeta-2\eta \gg 1$. 
As a result, the constraint $\zeta+\eta \gg1$ and the Eq. (\ref{eq14}) imply that
\begin{equation}
 \zeta+\eta = n +\frac{1}{2} \gg 1,
\end{equation}
or equivalently, 
\begin{equation}
n\gg 1.
\end{equation}
Consequently, the field variables can be approximated as
\begin{equation}
\zeta \simeq n  \gg 1;~ u \simeq \frac{3}{4} n^2 \gg 1; ~v\simeq \frac{9}{4}n \gg1
\end{equation}
during the inflationary phase with $n \gg 1$. 
It is also straightforward to show that $\eta <1/3$ and $\Sigma/H <1$ during the inflationary phase. 
For example, $\zeta \simeq 40.4974$, $\eta \simeq 0.00264$, and $\Sigma/H \simeq 0.000065$ if we choose $n=40$ and $\lambda=1$. 
For a comparison with the KSW model \cite{MW}, we will have $\zeta_{\rm KSW} = 40.1831$, $\eta_{\rm KSW}=0.316872$, and $(\Sigma/H)_{\rm KSW}=0.007886$ if we choose $\rho =40$ and $\lambda=1$ such that $n=\rho/\lambda$. It is clear that the spatial anisotropy parameter in the $J^2(X)F^2$ model is much smaller than that of the  KSW model with $f^2(\phi)F^2$. 
On the other hand the isotropy parameters are quite similar to each other. 

\section{Stability analysis of anisotropic solutions} \label{sec4}
In this section, we will try to show that the new set of power-law solutions are unstable during the inflationary phase. On the other hand, this new set of power-law solutions is stable during the slowly-expanding phase for a limited $\lambda$-$n$ domain.
\subsection{Inflationary phase}
In order to understand the stability of the set of power-law solutions, we will consider the power-law perturbation of the field equations with $\delta \alpha = A_{\alpha} t^{m}$, $\delta\sigma=A_{\sigma}t^{m}$, and $\delta\phi = A_{\phi}t^m$ ~\cite{WFK,WFK1}. 
Perturbing  Eqs. (\ref{eqnphi}), (\ref{eqnalpha}), and (\ref{eqsigma}) leads to a set of algebraic equations that can be cast as a matrix equation:
\begin{equation}\label{matrix}
{\cal D} \left( {\begin{array}{*{20}c}
   A_\alpha  \\
   A_\sigma  \\
   A_\phi \\
 \end{array} } \right) \equiv \left[ {\begin{array}{*{20}c}
   {A_{11} } & {A_{12} } & {A_{13} }   \\
   {A_{21} } & {A_{22} } & {A_{23} } \\
   {A_{31} } & {A_{32} } & {A_{33} }  \\
 \end{array} } \right]\left( {\begin{array}{*{20}c}
   A_\alpha  \\
   A_\sigma  \\
   A_\phi \\
 \end{array} } \right) = 0,
\end{equation}
with  
\begin{eqnarray}
 \label{a1j}
A_{11}=&&~\left({\lambda  n v-\frac{6}{\lambda}}\right)m-4\lambda  nv \left({\zeta+4\eta-4n-1}\right);~ A_{12}=4\lambda nv m-4\lambda  n v\left({\zeta+4\eta-4n-1}\right), \nonumber \\
A_{13}=&&~-\left[{\frac{\lambda^2vn}{2}(4n+1)-1}\right] m^2-\lambda^2 v\left[{8 n^3-2\left({\zeta+4\eta-2}\right) n^2 -\frac{1}{2} \left({\zeta+4\eta-1}\right)n-\frac{1}{\lambda^2v} \left(3\zeta-1\right)}\right] m +\lambda^2 u, \\
 \label{a2j} 
A_{21}=&&~m^2+\left({6\zeta-1}\right)m+\frac{2v}{3}\left({6n+1}\right); ~A_{22}=\frac{2v}{3}\left({6n+1}\right);~ A_{23}=-\frac{\lambda nv}{3} \left({6n+1}\right)m-\lambda u,\\
 \label{a3j}
A_{31}=&&~3\eta m +\frac{4}{3}v; ~A_{32}=m^2+\left({3\zeta-1}\right)m+\frac{4}{3}v;~ A_{33}=-\frac{2\lambda}{3}nv  m. 
\end{eqnarray}
Nontrivial solutions of Eq. (\ref{matrix}) exist only when
\begin{equation} \label{determinant}
\det {\cal D} =0.
\end{equation}
As a result, this determinant equation leads to a polynomial equation of $m$ :
\begin{equation} \label{poly}
m f(m) \equiv m\left({a_6 m^5+a_5m^4+a_4 m^3+a_3 m^2+a_2 m+a_1}\right)=0,
\end{equation}
with the coefficients $a_i$'s ($i=1-6$) given by
\begin{eqnarray}
a_6 &=&-\frac{1}{2}\left({4\lambda^2 v n^2 +\lambda^2 v n -2}\right),\\
a_5&=&-\lambda^2 vn \left[8n^2+8\left(2\zeta-\eta\right)n+\frac{1}{2} \left(8\zeta-4\eta-1 \right)\right] +3 \left(4\zeta-1\right),\\
a_4&=&-2\lambda^2 v \left(36\zeta+3v-8\right)n^3-\lambda^2 v \left(18\zeta^2 -72\zeta \eta +22\zeta+16\eta +3v-6+\frac{12}{\lambda^2}\right)n^2 \nonumber\\
&& -\frac{\lambda^2 v}{2} \left(9\zeta^2 -36\zeta \eta +2\zeta +8\eta +2v -1-\frac{4}{\lambda^2}\right)n +45\zeta^2 -24\zeta +\lambda^2 u +2v+3,\\
a_3&=& -2\lambda^2 v \left[72\zeta^2  +v\left(9\zeta +1\right)-36\zeta +4\right]n^3 \nonumber\\
&&+\lambda^2 v \left[18 \zeta^2\left(2\zeta-5\right)  +8\eta \left(18\zeta^2-9\zeta+1\right)  -v\left(7\zeta-14\eta+3\right)+38\zeta -4 -\frac{12}{\lambda^2} \left(3\zeta-1\right)\right]n^2 \nonumber\\
&&+
\frac {\lambda^2 v}{2}\left[9 \zeta^2 \left(2\zeta-3\right)+2\eta \left(36\zeta^2 -18\zeta+2\right)  -2v\left(4\zeta-5\eta \right) +10\zeta+2u-1+\frac{12}{\lambda^2} \left(3\zeta-2\eta-1\right) \right] n \nonumber\\
&&+3\zeta \left(18\zeta^2-15\zeta+4\right) +2v \left(8\zeta-\eta-2\right)+ \lambda^2 u \left(9\zeta-2-\frac{6}{\lambda^2} \right) -1,
\end{eqnarray}
\begin{eqnarray}
a_2&=&-8\lambda^2 v^2 \left(3\zeta-3\eta-1\right)n^3 +2\lambda^2 v^2 \left[3\zeta^2 +9\zeta \eta -12\eta^2 -14\zeta +\eta  +3+\frac{8u}{v}\right]n^2 \nonumber\\
&&+\lambda^2 v \left[v \left(5\zeta^2 +19 \zeta \eta -4\eta^2 -6\zeta -3\eta+1 \right) - u\left(\zeta +28\eta -7\right) +\frac{4}{\lambda^2} \left(3\zeta-1 \right)\left(3\zeta-3\eta-1\right)\right]n,\nonumber\\
&&+\lambda^2 u \left(18\zeta^2 -9\zeta+2v+1\right)+2\zeta v\left(15\zeta -3\eta -8\right) -6u \left(3\zeta-1\right)+2v\left(\eta+1\right),\\
a_1 &=&{2\lambda^2uv} \left[ { 8\left({3\zeta-3\eta-1}\right)n^2 -2\left({3\zeta^2+9\zeta \eta -12\eta^2-7\zeta + 2\eta +v +2}\right) n  +5\zeta-\eta-1 -\frac{4}{\lambda^2} }\right].
\end{eqnarray}

Positive root to Eq. (\ref{poly}) does not exist if all the coefficients $a_i$ are positive (or negative). 
As a result, the corresponding anisotropic solution is stable. 
On the other hand,  Eq.  (\ref{poly}) admits at least one positive root if $a_1 a_6 <0$ \cite{WFK,WFK1}.  
Therefore the corresponding anisotropic solution is unstable if $a_1 a_6 <0$. 

We can show that $a_6<0$ with the help of the inequality (\ref{inequality}). 
In addition,   the leading term of $a_1$ is given by  
\begin{equation}
a_1 \simeq {2\lambda^2uv}\left({18n^3+\frac{3}{2} n^2+n -\frac{4}{\lambda^2} }\right)
\end{equation}
for $n\gg 1$. 
Hence the inequality (\ref{inequality}) implies that $a_1>0$ for $n\gg 1$.  
This implies that the anisotropic power-law solution of the $J^2(X)F^2$ model is indeed unstable during the inflationary phase in contrast to the results of the KSW model.
The result is, however, consistent with the cosmic no-hair conjecture.  
It shows that the non-trivial coupling, $I\left({\phi,X,...}\right)F_{\mu \nu } F^{\mu \nu }$ is capable of inducing a small spatial-anisotropy for the evolution of our current universe. 

Note that the stability of the anisotropy depends on the choice of the function $I$ coupled to the $U_1$ gauge field. 
In particular, the anisotropy will survive the inflation if $I$ is chosen to be a canonical function of a scalar field as shown in the KSW model ~\cite{MW,WFK1}. 
On the other hand, if $I$ is chosen as a function of the kinetic term of the scalar field, the anisotropy is unstable during the inflationary phase.
As a result, the power-law solutions shows that our universe acts in favor of the forming of an isotropic (de Sitter) space consistent with that the cosmic no-hair conjecture predicts. 
\subsection{Slowly expanding phase}
Even the power-law solution is unstable during the inflationary phase, it will be interesting to know whether it is unstable during the slowly-expanding phase.
Indeed, we will show that the new set of power-law solutions is stable during the slowly-expanding phase for a small $\lambda$-$n$ domain.
Note that $\zeta+\eta $ and $\zeta-2\eta$ are of order one, ${\cal O}(1)$, for the 
slowly expanding phase. 
Note again that unstable solution will not exist if all coefficients $a_i$ listed above are positive (or negative) definite. 
In fact, the equation $f(m)=0$ admits no positive root for a wider range of $\lambda$ and $n$. 

\begin{figure}[hbtp]
\begin{center}
{\includegraphics[height=70mm]{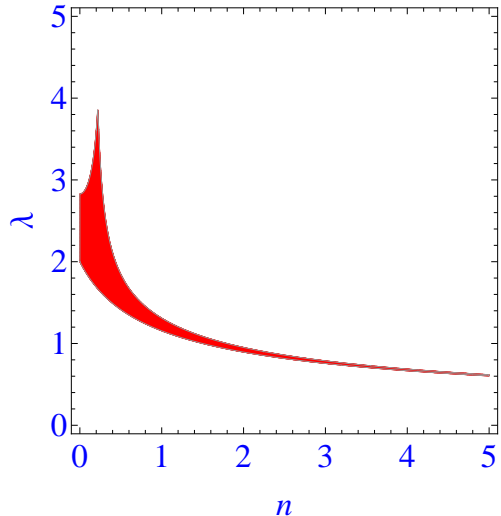}}\\
\caption{The red region represents the $\lambda$-$n$ domain for the existence of stable slowly expanding solutions.} 
\label{fig1}
\end{center}
\end{figure}

Indeed,  we will plot a $\lambda$-$n$ domain numerically in which all real parts of the five roots of $f(m)=0$ are non-positive, i.e., ${\rm Re}(m_i)\leq 0$, during the slowly expanding phase.  The result is shown in Fig. \ref{fig1}.

According to the result shown in Fig. \ref{fig1}, this allowed domain is quite limited. 
In particular, the range of $\lambda$ is much wider for small $0<n\leq1$.
Here we have set $\zeta+\eta >0$, $\zeta-2\eta >0$, $u>0$, and $v>0$ for expanding solutions. 
Note again that we will only have unstable anisotropic inflationary solutions for the large $n$ as proved earlier in this section.

In addition, we would like to see whether the new set of slowly expanding solutions is indeed a set of attractive solutions for the parameters $\lambda$ and $n$ chosen within the red region in Fig. \ref{fig1}. 
Following Refs. \cite{MW,WFK1},  we will introduce the dynamical variables
\begin{equation}
{\hat X}= \frac{\dot\sigma}{\dot\alpha};~{ Y} =\frac{\dot\phi}{\dot\alpha};~{ Z}= \frac{J_0^{-1}p_A}{\dot\alpha} \left(\frac{\dot\phi^2}{2}\right)^{-n}\exp\left[-2\alpha -2\sigma\right].
\end{equation}
As a result, a set of autonomous equations of the dynamical system can be derived from the field equations (\ref{eqnphi}), (\ref{eqnalpha}), and (\ref{eqsigma}) as
\begin{eqnarray}
\label{eqX}
\frac{d {\hat X}}{d\alpha} &=&  -\frac{\ddot\alpha}{\dot\alpha^2}{\hat X}+\frac{\ddot\sigma}{\dot\alpha^2}\nonumber\\
&=&{\hat X} \left[3 \left({\hat X}^2-1\right) +\frac{Y^2}{2} + \left(n+\frac{1}{3}\right) Z^2\right] +\frac{Z^2}{3},\\
\label{eqY}
\frac{d Y}{d\alpha}&=& -\frac{\ddot\alpha}{\dot\alpha^2}Y+\frac{\ddot\phi}{\dot\alpha^2}\nonumber\\
&=&Y \left\{ 3{\hat X}^2 +\frac{Y^2}{2} + \left(n+\frac{1}{3}\right)Z^2\right.\nonumber\\
&&\left. +\frac{1}{Y^2-2n(4n+1)Z^2} \left\{\lambda Y \left[3 \left({\hat X}^2-1\right)+\frac{Y^2}{2}+\left(2n+\frac{1}{2}\right)Z^2\right] +2n\left(4{\hat X}+1\right)Z^2 -3Y^2\right\} \right\},\\
\label{eqZ}
\frac{dZ}{d\alpha} &=& -Z\left[\frac{\ddot\alpha}{\dot\alpha^2}+2{\hat X}+2+\frac{2n}{Y}\frac{\ddot\phi}{\dot\alpha^2}\right] \nonumber\\
&=&Z \left\{ 3{\hat X}^2+\frac{Y^2}{2}+\left(n+\frac{1}{3}\right)Z^2 -2{\hat X}-2 \right. \nonumber\\
&& \left. -\frac{2n}{Y^2-2n(4n+1)Z^2} \left\{\lambda Y \left[3 \left({\hat X}^2-1\right)+\frac{Y^2}{2}+\left(2n+\frac{1}{2}\right)Z^2\right] +2n\left(4{\hat X}+1\right)Z^2 -3Y^2\right\} \right\}.
\end{eqnarray}
Note that we have used the Hamiltonian equation (\ref{friedmann}) 
\begin{equation} \label{dyn.constraint}
\frac{V}{\dot\alpha^2} =-3\left({\hat X}^2-1\right) -\frac{Y^2}{2} -\left(2n+\frac{1}{2}\right)Z^2,
\end{equation}
in order to derive the above equations. 
Note that the anisotropic fixed points are solutions of the autonomous equations $d{\hat X}/d\alpha =dY/d\alpha=dZ/d\alpha=0$. 
As a result, we can obtain, from equations $dY/d\alpha=dZ/d\alpha=0$, a relation given by
\begin{equation} \label{dyn.eq1}
3 {\hat X}^2 +\frac{Y^2}{2} + \left(n+\frac{1}{3}\right) Z^2 =\frac{2({\hat X}+1)}{2n+1} .
\end{equation}
The equation $d{\hat X}/d\alpha=0$ can thus be reduced to
\begin{equation}\label{dyn.eq2}
Z^2 =-3 {\hat X} \left[\frac{2({\hat X}+1)}{2n+1} -3\right].
\end{equation}
On the other hand,  the equations $d{\hat X}/d\alpha =dY/d\alpha=dZ/d\alpha=0$ will lead the Hamiltonian constraint (\ref{dyn.constraint}) given by
\begin{equation}
\frac{d}{d\alpha} \left(\frac{V}{\dot\alpha^2} \right) =0.
\end{equation}
Hence we have 
\begin{equation}
\frac{\ddot\alpha}{\dot\alpha^2} =\frac{\lambda}{2}Y,
\end{equation} 
or equivalently,
\begin{equation} \label{dyn.eq3}
3 {\hat X}^2 +\frac{Y^2}{2} + \left(n+\frac{1}{3}\right) Z^2 =-\frac{\lambda}{2}Y.
\end{equation}
As a result, the relation between ${\hat X}$ and $Y$ can be shown to be
\begin{equation} \label{dyn.eq4}
Y= -\frac{4\left({\hat X}+1\right)}{\lambda(2n+1)} 
\end{equation}
from Eqs. (\ref{dyn.eq1}) and (\ref{dyn.eq3}).
With the useful relations shown in Eqs. (\ref{dyn.eq2}) and (\ref{dyn.eq4}), we are able to rewrite Eq. (\ref{dyn.eq1}) as an equation of ${\hat X}$ given by
\begin{equation}
\left[\lambda^2\left(2n+1\right)+8\right] {\hat X}^2 + \left[\lambda^2 \left(36n^3 +36n^2+7n-1\right)+16\right]{\hat X}-2\lambda^2 \left(2n+1\right)+8=0.
\end{equation}
Consequently,  non-trivial solutions of ${\hat X}$ can be solved as
\begin{equation}
{\hat X}_\pm =-\frac{\lambda^2 \left(2n+1\right) \left(18n^2+9n-1\right) +16 \pm 3 \lambda\left(2n+1 \right) \sqrt{\lambda^2\left({36n^4+36n^3+5n^2-2n+1}\right)+32n}}{2 \left[\lambda^2(2n+1)+8\right]}.
\end{equation}
It is apparent that the ${\hat X}_{-}$ is identical to the ratio $\Sigma/H$ of the anisotropic power-law solution found in the previous section. 
In other words, ${\hat X}_{-}$ is exactly the anisotropic fixed point equivalent to the anisotropic power-law solution we found earlier.
 
\begin{figure}[hbtp]
\begin{center}
{\includegraphics[height=70mm]{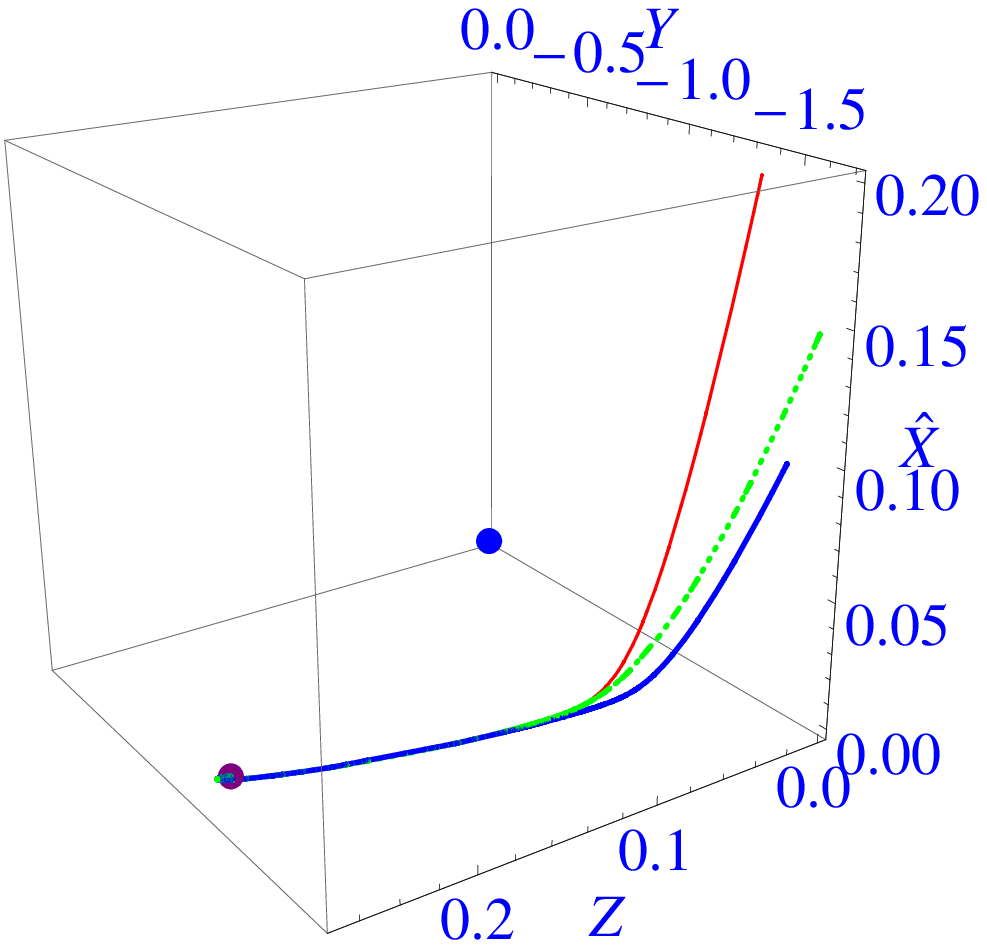}}\quad
{\includegraphics[height=70mm]{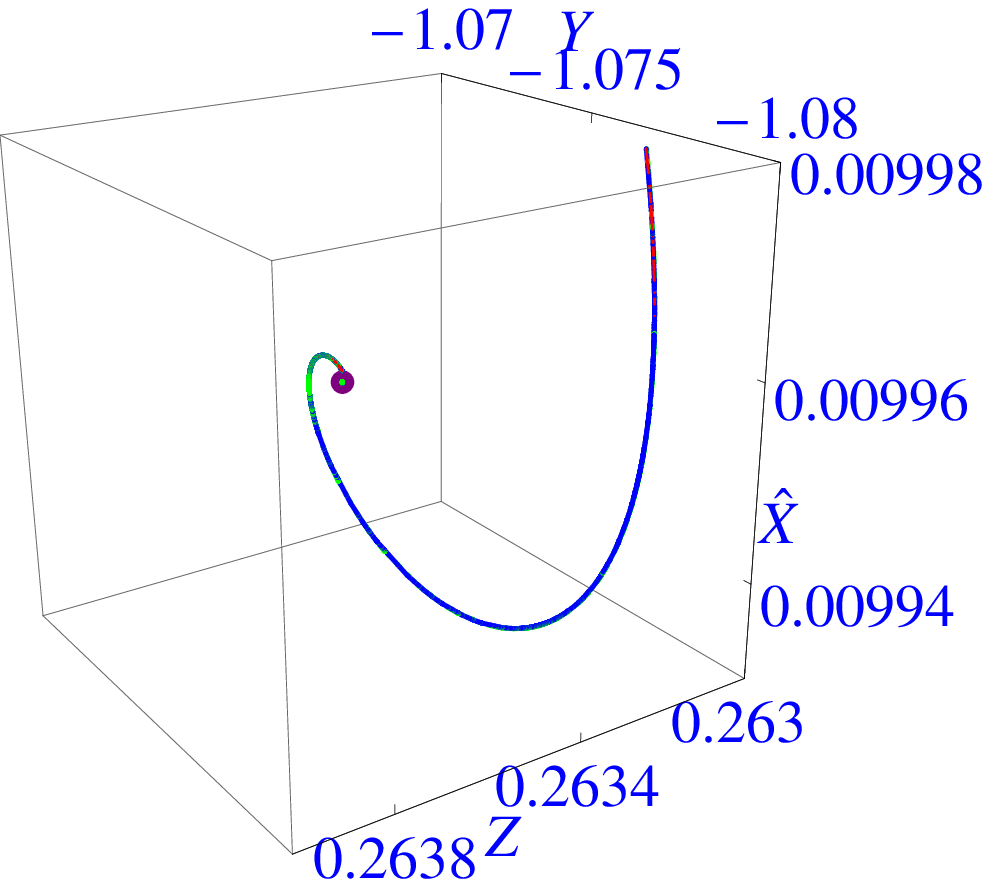}}
\caption{The anisotropic fixed point with the field parameters chosen as $\lambda=1.25$ and $n = 1$ is shown as an attractor solution in this figure. 
All trajectories in the phase space of $\hat X$, $Y$, and $Z$, with different initial conditions converge to  the anisotropic fixed point (plotted as the purple point) rather than the isotropic fixed point $({\hat X},~Y,~Z)=(0,~0,~0)$ (plotted as the blue point). 
The initial conditions $({\hat X}(t=0),~Y(t=0),~Z(t=0))$ are set as $(0.2,~ -1.5,~ 0.02)$ for the thin-solid red curve, $(0.1,~-1.6,~0.01)$ for the thick-solid  blue  curve, and $(0.15,~-1.7,~0.01$)  for the green dotted curve, respectively.
The right figure is the enlarged picture of the left figure.} 
\label{fig2}
\end{center}
\end{figure}

Hinted by the allowed region of $\lambda$ and $n$ for the existence of stable expanding-solutions shown above, we can, for example, examine the attractor behavior of the anisotropic fixed point for $n=1$ (corresponding to the canonical kinetic term) and $\lambda=1.25$. 
Note that for this choice the scale factors will be $\zeta \simeq 1.5$ and $\eta \simeq 0.015 \ll \zeta$. The result is shown in Fig. \ref{fig2}.

As a result, the anisotropic fixed point can be shown to be the attractor to the dynamical system as expected. 
This result shows that the $J^2(X)F^2$ model does admit stable and attractive anisotropic expanding solutions even this model does not admit any stable and attractive anisotropic inflationary solution. 
Nevertheless, the cosmic no-hair conjecture is indeed violated in the $J^2(X)F^2$ model. 
A small anisotropy will still sustain during the slowly-expanding phase and leads to a small anisotropy observed today.

\section{Conclusions} \label{sec5}
Inflation has been considered as a main paradigm in the current cosmology. 
It is successful in solving some fundamental questions in cosmology.
The predictions are also consistent with the observations of the cosmic microwave background radiation. 
In addition, the cosmic no-hair conjecture explains the physical origin of the highly isotropic universe. 
This conjecture is based on a belief that the vacuum energy dominance would erase any classical hair. The conformal-violating Maxwell model  ~~\cite{ITD,MST,BR,DMR,Grasso:2000wj,Barrow:2011ic} proposed in Ref. ~\cite{MW} provides a counter-example to the cosmic no-hair theorem.
Hence we  propose to study a conformal-violating Maxwell model with a coupling term, $J^2(X)F^2=J_0^2 X^{2n}  F^2$. 

In contrast to results found in ~\cite{MW,WFK1}, we found that the new model does not  admit any stable and attractive Bianchi type I inflationary solution. 
By a careful stability analysis, we have found, however, that the proposed model does admit stable and attractive Bianchi type I expanding solutions in the slowly-expanding phase. 
Hence this model provides another counter-example to the cosmic no-hair conjecture with a tiny spatial hair.
The results shown in this paper suggest that a correlation between an anisotropy of universe space and a broken conformal invariance deserve more attentions.
 
Moreover, small anisotropy is required to explain  the origin of  large-scale galactic electromagnetic field in the present universe.
It would be interesting to examine the validity of the cosmic no-hair conjecture for some  other possible coupling terms in the conformal-violating theory. For example, those models that have been discussed recently, e.g., $ I\left(R,R_{\mu\nu},R_{\mu\nu\lambda\kappa}\right)$ ~\cite{ITD,MST,Adak:2016led,FDM}, $I=I\left({A^2}\right)$ ~\cite{AG}, $I=I\left({G}\right)$ ~\cite{MRS,GEFC}, and $I=I\left[{(k_F)_{\alpha\beta\mu\nu}}\right]$ ~\cite{LCVAK}. Hopefully, the result shown here could be helpful to the study on the cosmic evolution of our universe. 

{\bf Note added}: 
The appearance of Ref. \cite{Holland:2017cza} came to our attention when are finalizing the present paper. Motivated by Ref. \cite{Dimopoulos:2011pe}, Ref. \cite{Holland:2017cza} proposes an extension of the KSW model with $f(\phi) \to f(\phi, X)$ similar to ours.
In particular,  they have pointed out that no small-anisotropy inflationary solutions exist for $f(X) \propto X^{-n}$ in contrast to our  model with $f(X) \propto X^n$. 
\begin{acknowledgments}
T.Q.D. is supported in part by the Vietnam National Foundation for Science and Technology Development (NAFOSTED) under grant number 103.01-2017.12. W.F. Kao is supported in part by the Ministry of Science and Technology (MOST) of Taiwan under Contract No. MOST 104-2112-M-009-020-MY3.
 \end{acknowledgments}



\end{document}